\documentclass[twocolumn,trackchanges]{aastex7}

\usepackage{amsmath}

\begin{document}

\title{SMA observations of the archetypal CSS radio source 3C 303.1}

\author[orcid=0000-0003-4506-1352,gname=Rushikesh,sname=Bhutkar]{Rushikesh Bhutkar}
\affiliation{Department of Physics and Astronomy, University of Manitoba, Winnipeg, MB R3T 2N2, Canada}
\email[show]{bhutkarr@myumanitoba.ca}  

\author[orcid=0000-0003-0685-3621,gname=Mark,sname=Gurwell]{Mark Gurwell}
\affiliation{Center for Astrophysics \textbar\ Harvard \& Smithsonian, 60 Garden Street, Cambridge, MA 02138, USA}
\email{mojegan.azadi@cfa.harvard.edu}

\author[orcid=0000-0001-6421-054X,gname=Christopher,sname=O'Dea]{Christopher O’Dea}
\affiliation{Department of Physics and Astronomy, University of Manitoba, Winnipeg, MB R3T 2N2, Canada}
\affiliation{Center for Space Plasma \& Aeronomic Research 
University of Alabama in Huntsville
Huntsville, AL 35899, USA}
\email{Christopher.O'Dea@umanitoba.ca}

\author[orcid=0000-0001-6004-9728,gname=Mojegan,sname=Azadi]{Mojegan Azadi}
\affiliation{Center for Astrophysics \textbar\ Harvard \& Smithsonian, 60 Garden Street, Cambridge, MA 02138, USA}
\email{}

\author[orcid=0000-0003-1809-2364,gname=Belinda,sname=Wilkes]{Belinda Wilkes}
\affiliation{Center for Astrophysics \textbar\ Harvard \& Smithsonian, 60 Garden Street, Cambridge, MA 02138, USA}
\affiliation{H.H. Wills Physics Laboratory, University of Bristol, Tyndall Avenue, Bristol, BS8 1TL, UK}
\email{bwilkes@cfa.harvard.edu}

\author[orcid=0000-0002-4735-8224,gname=Stefi,sname=Baum]{Stefi Baum}
\affiliation{Department of Physics and Astronomy, University of Manitoba, Winnipeg, MB R3T 2N2, Canada}
\affiliation{Center for Space Plasma \& Aeronomic Research 
University of Alabama in Huntsville
Huntsville, AL 35899, USA}
\email{Stefi.Baum@umanitoba.ca}

\author[orcid=0000-0002-4464-8023,gname=Dhruba,sname=Saikia]{D. J. Saikia}
\affiliation{Fakultat f\"ur Physik, Universit\"at Bielefeld, Postfach 100131, D-33501 Bielefeld, Germany}
\affiliation{Assam Don Bosco University, Guwahati 781017, Assam, India}
\email{dhrubasaikia.tifr.ccsu@gmail.com}

\author[orcid=0000-0002-5445-5401,gname=Grant,sname=Tremblay]{Grant Tremblay}
\affiliation{Center for Astrophysics \textbar\ Harvard \& Smithsonian, 60 Garden Street, Cambridge, MA 02138, USA}
\email{granttremblay@gmail.com}

\begin{abstract}

AGN (active galactic nuclei) feedback plays a crucial role in shaping galaxy evolution. Numerical simulations show that the kinetic energy transfer efficiency from radio jets to interstellar medium (ISM) is relatively high, suggesting a significant role of radio jets in the feedback. We investigate AGN feedback on cold gas in the compact steep-spectrum (CSS) radio source 3C 303.1. CSS sources are largely young radio galaxies evolving through dense environments of their host galaxies. This early evolutionary phase likely represents a critical stage in which the radio source has maximum impact on host galaxy evolution. 3C 303.1 is the only CSS source so far showing alignment in optical, X-ray and UV with the radio structure, making it an interesting source to investigate jet-feedback. We present continuum and spectral line observations of the J=3-2 transition of carbon monoxide ($^{12}$CO) of 3C 303.1, obtained with the Submillimeter Array. We detect continuum emission at 221.1 and 271.2~GHz. We do not detect the $^{12}$CO(J=3-2) line and derive an upper limit on the molecular gas mass of $\sim 2.3 \times 10^{9}\ \mathrm{M_\odot}$. The gas-to-dust mass (G/D) ratio is found to be at the lower end of typical Galactic values, and the star-formation rate (SFR) derived is moderate but declining likely due to a recent  quenching event. Therefore, $^{12}$CO(J=3-2) non-detection, relatively low G/D ratio, and the moderate but declining SFR point to the shock-heating and/or removal of CO gas by AGN, consistent with the AGN feedback signatures observed in 3C 303.1.
\end{abstract}


\section{Introduction} \label{sec:1}

Almost all massive galaxies host supermassive black holes (SMBHs) with masses $ \sim 10^7-10^9 \,
\mathrm{M_\odot} $ at their centers. Sometimes, these black holes go through the phase where they actively accrete material from their surroundings, producing enormous amounts of energy across the electromagnetic spectrum; this is known as an active galactic nucleus (AGN; \cite{Rees1984, Urry1995, Padovani2017a}). An AGN injects energy into the host galaxy and its surrounding medium via winds, radiation, and relativistic jets. This process, known as AGN feedback, is thought to maintain the observed tight correlation between central SMBH mass and stellar velocity dispersion ($\sigma$) in the bulge of the host galaxy by regulating the gas reservoir of the galaxy \citep{Morganti2017, Harrison2024, 1998A&A...331L...1S}. In order to match observations, all cosmological models now include AGN feedback, as it solves some of the long-standing problems such as the cooling flow problem in the centers of the cool-core cluster galaxies--where hot gas is expected to radiatively cool, creating more stars, and the discrepancy at the brighter end of the luminosity function of the galaxies, where models overpredict the number of massive galaxies \citep{10.1093/mnras/250.4.737, 2007ARA&A..45..117M, 2006MNRAS.365...11C}. In both cases, energy injected by the AGN feedback heats or ejects the gas from the galaxy's surroundings, suppressing both star formation and further accretion onto the SMBH. AGN feedback plays a crucial role in shaping galaxy evolution and demonstrates the coevolutionary relationship of central SMBHs and their host galaxies.

The classical AGN paradigm consisted of two complementary modes: radio mode feedback (also known as maintenance/kinetic mode), driven by mechanical jets and quasar mode feedback (also known as radiative mode), driven by radiation and winds. Quasar mode feedback is radiatively efficient, characterized by a high accretion rate onto the SMBH, and affects the cold gas \citep{Haehnelt1998}. In contrast, radio mode feedback is radiatively inefficient, with a low accretion rate, where relativistic jets inflate cavities in the hot intracluster medium (ICM) \citep{Fabian2012, 2006MNRAS.365...11C, Mcnamara2012}. This paradigm represented an important step in understanding the coevolution of SMBHs and their host galaxies. Now there is growing evidence of radio sources having a substantial effect on the cold component of the interstellar medium (ISM): 1) Cold molecular gas is being driven by radio jets in massive galactic fountains observed in cool-core, brightest cluster galaxies (BCGs) \citep{2018ApJ...865...13T}. 2) Radio jets appear to drive shocks that push out neutral atomic hydrogen (H\textsc{i}) and molecular gas, suggesting jet-ISM interactions \citep{refId0,morganti2013radio}. 3) Shocks generated by jets can heat the gas, thereby delaying or suppressing star formation, an effect referred to as negative feedback \citep{2007ARA&A..45..117M, Lanz2016}. 4) Conversely, these shocks may also compress molecular gas, indirectly enhancing the star formation locally, representing positive feedback \citep{Labiano2008a, Duggal2024, Mukherjee2018b}.

This evidence has shown that the current feedback paradigm is too basic and requires revision. Additionally, current cosmological models omit jet feedback in their simulations because the physical nature of jet-ISM interactions remains poorly understood \citep{10.1093/mnras/stae1021, 10.1093/mnras/stad2640, 10.1093/mnras/staa1894}. Jets carry kinetic energy approximately an order of magnitude greater than that of AGN winds \citep{2023Galax..11...21H}. Numerical simulations show that the kinetic energy transfer efficiency from radio jets to ISM is relatively high, ~20-30\%, suggesting a significant role of radio jets in the feedback \citep{2011ApJ...728...29W, 2016MNRAS.461..967M}. In addition, the coupling between radio jet power and outflow kinetic power is critical but poorly understood. Therefore, it is essential to study how radio sources interact with cold gas in galaxies to inform theoretical models to produce a more accurate and revised version of the AGN feedback.

Compact steep-spectrum (CSS) radio sources are among the best examples of radio sources that drive multiphase gas outflows \citep{10.1111/j.1365-2966.2012.21247.x,refId0,morganti2013radio}. CSS sources are defined as having total projected linear sizes between 1 kpc and 20 kpc, with a steep synchrotron spectrum. ($\alpha > 0.5$,  where $S(\nu) \propto \nu^{-\alpha}$) (see review \cite{2021A&ARv..29....3O}). They are believed, in many cases, to be young objects that advance through dense inhomogeneous and asymmetric environments of their host galaxies as they journey toward evolving into larger radio sources \citep{1995A&A...302..317F, 1998PASP..110..493O}. Their sub-galactic sizes make them excellent laboratories for studying jet-ISM interactions. They interact vigorously with the ISM of the host galaxy by driving outflows of the multiphase gas. As these radio sources propagate through the ISM, they drive shocks at velocities of thousands of km/s, which is similar to a (scaled-up) supernova blast wave \citep{ODea2002}. This early evolutionary phase likely represents a critical stage in which the evolution of the radio galaxy has a maximum impact on the host galaxy \citep{Best2000, debreuck2000}. There is substantial observational evidence that these compact, powerful radio sources strongly interact with the ISM of their host galaxies \citep{2021A&ARv..29....3O}. Therefore, CSS sources are key systems for understanding jet-ISM interactions.

In this paper, we investigate radio-mode feedback on cold molecular gas in the CSS radio galaxy 3C 303.1 (z = 0.267). We use carbon monoxide ($^{12}$CO) as a tracer for the cold molecular gas \citep{1998Natur.395..871N}. The goal is to detect $^{12}$CO and search for bipolar outflows; therefore, high-resolution observations are needed. Due to its northern declination, 3C 303.1 is inaccessible to the Atacama Large Millimeter/submillimeter Array (ALMA); therefore, we observed 3C 303.1 using the Submillimeter Array (SMA) in Maunakea, Hawaii. We target the J=3-2 transition of $^{12}$CO at 345.8 GHz redshifted to 272.928 GHz, which is the only $^{12}$CO transition that falls within the SMA frequency coverage. There are no previous observations of the $^{12}$CO(J=1-0) and $^{12}$CO(J=2-1).
In AGN hosts, intense X-ray heating maintains higher gas temperatures, which excite CO molecules to higher rotational (J) levels. Such high transitions are produced in denser ($n_{H_2} = 10^5-10^6 \mathrm{cm}^{-3}$) and warmer ($T_{k} = 50-600\,\mathrm{K}$) gas. Hence, higher-J transitions are brighter up to J=7-6 than low-J transitions \citep{Esposito2024}. The $^{12}$CO(J=3-2) line is therefore expected to be strong compared to $^{12}$CO(J=1-0) and can provide meaningful upper limits. We obtained both continuum and $^{12}$CO(J=3-2) observations at 221.1 GHz and 271.2 GHz respectively. 

The paper is organized as follows. Section \ref{sec:2} presents our motivation for studying 3C 303.1. In Section \ref{sec:3}, we describe the SMA data reduction and preparation of the spectral energy distribution (SED), including archival photometry and extinction corrections. In section \ref{sec:4}, we discuss the discrepancy between our 221.1 GHz and the previously reported 230 GHz flux-density measurement from the literature. We then derive upper limits on the $^{12}$CO(J=3-2) line intensity, as well as on the molecular gas mass. We also describe the SED fitting procedure across the radio-to-X-ray range and discuss the output parameters from SED fitting. Our findings are summarized in Section \ref{sec:5}.

Throughout this paper, we adopt H$_0$= 67.66 km s{$^{-1}$} Mpc {$^{-1}$}, $\Omega _m$ = 0.31 , $\Omega_\Lambda = 0.69$ \citep{Planck2018}. The spatial scale is 4.2 kpc/$^{\prime\prime}$. 

\section{3C 303.1} \label{sec:2}

\begin{figure}[ht!]
\plotone{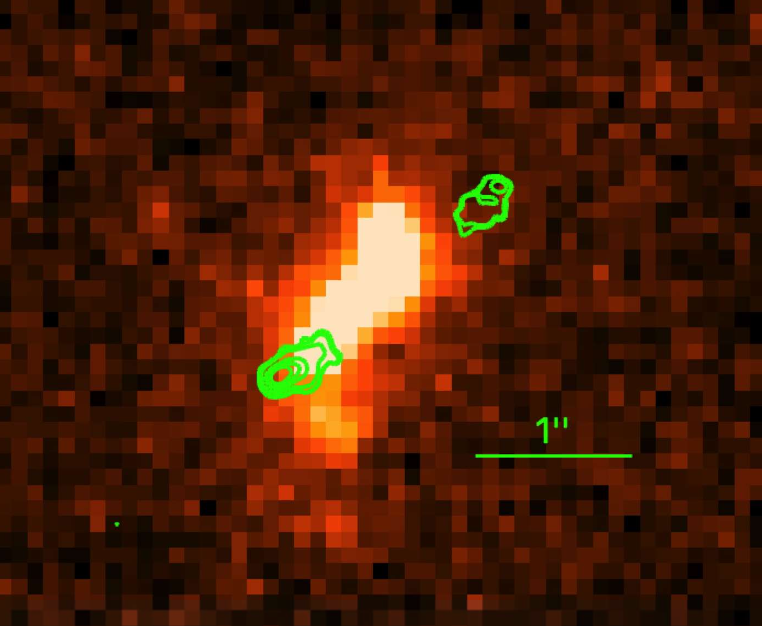}
\caption{HST/WFPC2/LRF image showing continuum-subtracted [O$\,\textsc{iii}]$ emission \citep{deVries1999} on the 5 GHz MERLIN contours \citep{Ludke1998}}
\label{fig:3C3303.1_OIII}
\end{figure}

3C 303.1 is a powerful and young CSS radio source.  High-resolution imaging reveals a double-lobed radio structure with the largest angular size of 1.9 arcseconds, corresponding to a projected linear size of 8 kpc \citep{Ludke1998}. Its optical spectrum shows narrow, high-excitation emission lines, and the source is classified as High-Excitation Radio Galaxy (HERG) \citep{Buttiglione2010b}. HERGs are characterized by radiatively efficient accretion disks with high accretion rates, typically  $\geq 1\%$ of the Eddington ratio \citep{Best2012, Best2014}. 

\begin{deluxetable*}{lr}
\tablecaption{Source properties \label{tab:properties}}
\tablehead{
\colhead{Parameter} & \colhead{3C 303.1}
}
\startdata
Optical classification & G \\
redshift (z) & $0.267^{\dagger}$ \\
RA (J2000) & 14$^{\text{h}}$43$^{\text{m}}$14\fs56 \\
Dec (J2000) & +77\degr07\arcmin27\farcs71 \\
Optical spectroscopic classification $^{1}$  & HERG \\
$\log_{10} \mathrm{Radio\,Power}\,(P_{5\,\mathrm{GHz}})$ W Hz$^{-1}$ & $25.99 \pm 0.01$ \\
Luminosity distance $D_{L}$ & 1402.5 Mpc \\
Observed $\text{H}\alpha$ luminosity $^{2}$&  $(1.26 \pm 0.01) \times 10^{42}$ \\ 
($L(\text{H}\alpha)_{\text{obs}}$) & $\text{erg\,s}^{-1}$ \\ 
Integrated emission line  $^{3}$ & $(6.59 \pm 0.07)\times 10^{42}$ \\
$L([O\,\textsc{iii}]~\lambda5007)$ & $\text{erg\,s}^{-1}$  \\
Integrated [O$\,\textsc{iii}]~\lambda5007$ FWHM $^{3}$ & $815 \pm 18 $ $\text{km\,s}^{-1}$\\
Spectral age of the radio emission$^{4}$ & $\sim10^5$ yr\\
\enddata
\tablecomments{References: (1)~\cite{Buttiglione2010b}; (2)~\cite{Buttiglione2009}; (3)~\cite{Gelderman1994}; (4)~\cite{ODea2002} \\
$^{\dagger}$ An alternative redshift of $z = 0.270$ is also reported in the literature \citep{Holt2008}; however, the difference is negligible and does not affect our results.}
\end{deluxetable*}

\begin{figure*}[ht!]
\gridline{
\fig{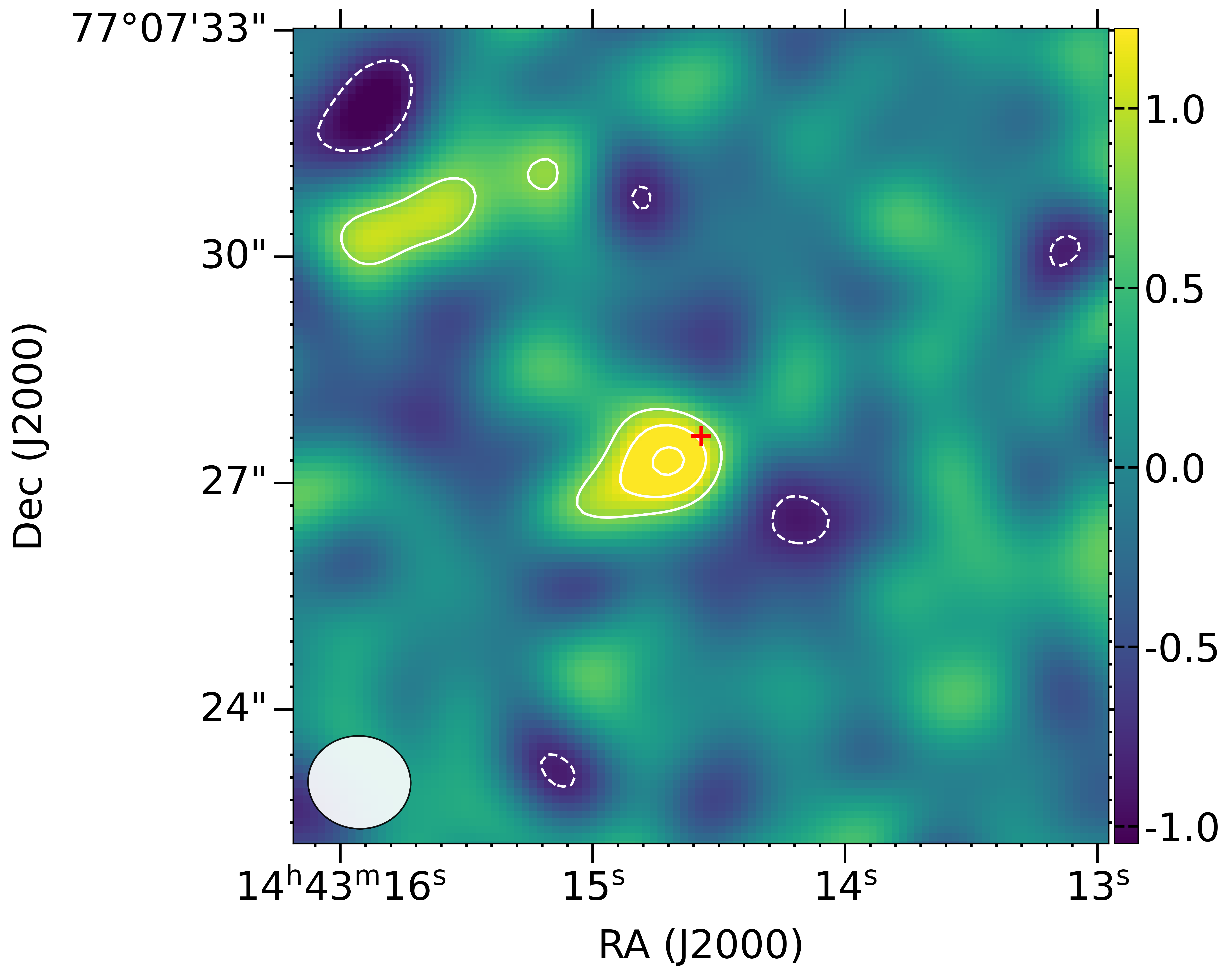}{0.45\textwidth}{(a)}
\fig{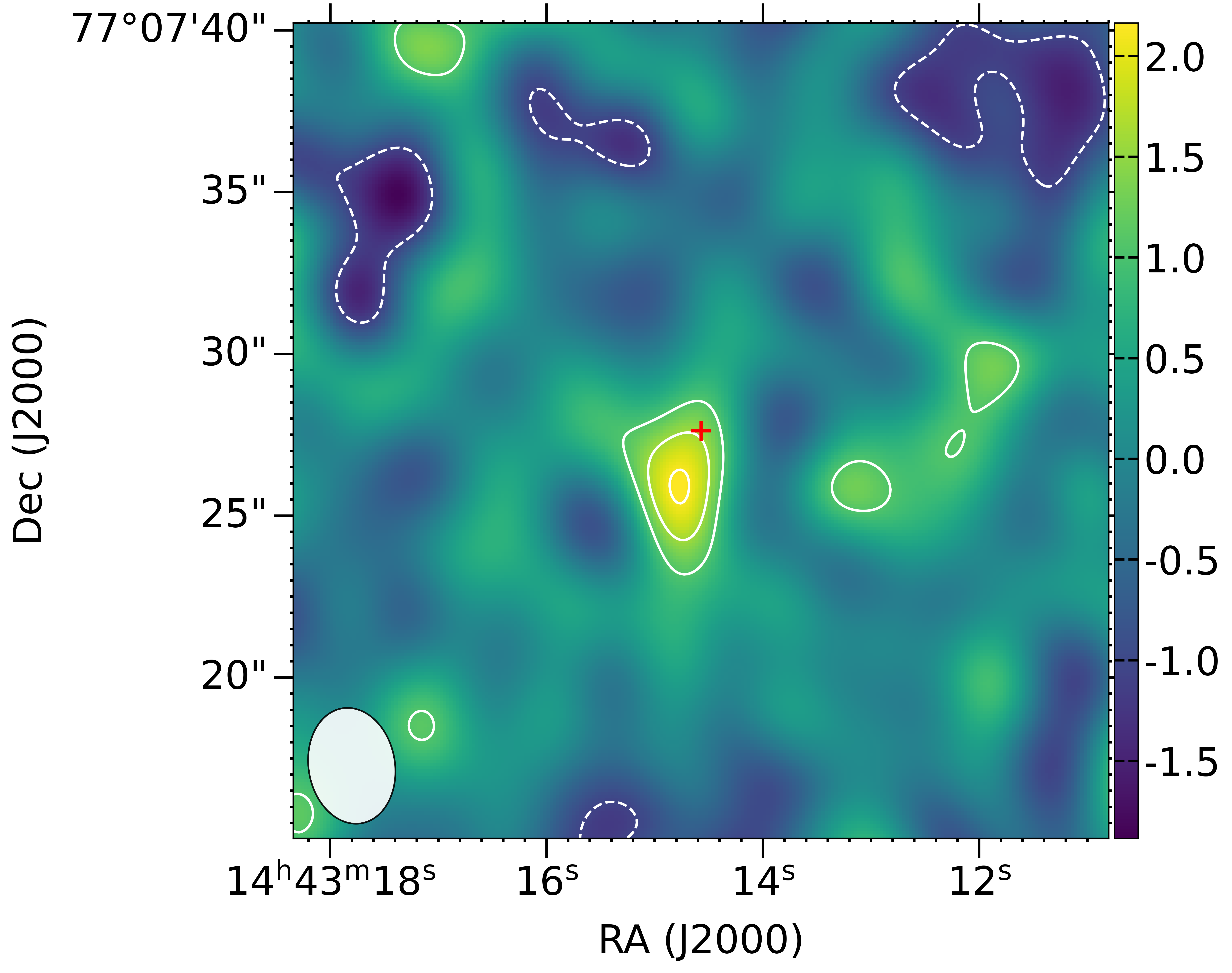}{0.45\textwidth}{(b)}
}

\caption{(a) Total intensity continuum map of 3C 303.1 at 221.1 GHz. The continuum is detected at SNR of 6.80. RMS noise ($\sigma$) in the map is 0.40 mJy beam$^{-1}$. The synthesized beam is $1.36^{\prime\prime} \times 1.23^{\prime\prime}$ with a major axis of position angle (P.A.) = 85.9$^\circ$. (see Table \ref{tab:observations_parameters}). (b) Total intensity continuum map of 3C 303.1 at 271.2 GHz. The continuum is detected at SNR of 3.83. $\sigma$ in the map is 0.53 mJy beam$^{-1}$. The synthesized beam is $3.60^{\prime\prime} \times 2.67^{\prime\prime}$ with a major axis of P.A. = 9.4$^\circ$. The contour levels in both maps correspond to $\sigma$ $\times$ [-2, 2, 3, 4] mJy beam$^{-1}$. The color scale shows flux density in mJy beam$^{-1}$. }
\label{fig:continuum_maps}
\end{figure*}

The spatial alignment of the optical emission line gas with the radio jet axis suggests that radio jets are driving outflows of the warm-ionized gas, consistent with the observed emission line kinematics (see Fig.\,\ref{fig:3C3303.1_OIII}) \citep{deVries1999, Axon2000, Shih2013}. \cite{Gelderman1994} measured velocities from the entire [O$\,\textsc{iii}]~\lambda5007$ emission-line profile after continuum subtraction and removal of the [O$\,\textsc{iii}]~\lambda4959$ and Fe$\,\textsc{ii}$ contributions. The line fluxes and width were measured by integrating the entire line profile. The line centre was measured from median (centre of area). The [O$\,\textsc{iii}]~\lambda5007$ Full Width at Half Maximum (FWHM) is 815 $\pm$ 18 km s$^{-1}$. \cite{Shih2013} used [O$\,\textsc{ii
}]$/[O$\,\textsc{iii}]$ vs.\,[O$\,\textsc{iii}]$/H$\beta$ line ratios to distinguish AGN ionization from shock ionization. The integrated [O$\,\textsc{iii}]~\lambda5007$ component, the velocity gradient aligned with the radio axis, and the line ratios consistent with shock-dominated ionization in the southeastern lobe, together indicate that the radio jets are driving strong outflows in 3C 303.1. \citep{Labiano2005, ODea2002, Shih2013}. The observed cloud velocities ($\sim300-500$ km s$^{-1}$) in 3C 303.1 are also consistent with moderate shock heating by bow shocks associated with expanding radio lobes, which could also contribute to the observed optical emission-line properties \citep{ODea2002}. X-ray and radio alignment seen in the \textit{Chandra} data is consistent with numerical simulations of X-ray emission produced by a hot and shocked gas \citep{Massaro2010, Heinz1998, Bicknell2006}. Observations from XMM-Newton also reveal a component that could originate from a hot, shocked gas \citep{ODea2006}. Another possibility for this X-ray emission is inverse-Compton scattering, arising from the southeastern lobe. HST/UV imaging shows two patches of UV emission along the radio axis, one of which coincides with the brighter southeastern lobe. The UV light may be interpreted as jet-induced star formation, i.e., positive feedback \citep{Labiano2008a}. 

The southeastern lobe has a maximum polarization of only about 1\%, compared to up to about 20\% for the northwestern lobe \citep{Ludke1998}. This is consistent with Faraday depolarization, suggesting that the southeastern side jet is interacting more strongly with the ambient medium, which it heats and ionizes. The southeastern lobe is brighter and closer to the nucleus is also a sign of interaction with the dense gas \citep{Saikia2003}. Only two other CSS sources (3C 237 and 3C 305) show aligned X-ray emission with the radio source structure \citep{Massaro2018, Massaro2010,10.1111/j.1365-2966.2012.21247.x}. So far, we know that seven CSS sources exhibit UV alignment with the radio sources \citep{Labiano2008a, Duggal2024}. Notably, 3C 303.1 is the only CSS source so far showing alignment in both X-ray and UV images with the radio structure. The spatial conincidence of optical, X-ray and UV emission with the radio source reveal that 3C 303.1 is driving shocks into the ISM of the host galaxy, making it an important low-redshift analogue for studying radio-mode feedback.

\begin{deluxetable*}{ccccccccc}
\tablewidth{0pt}
\tablecaption{SMA Observational Parameters \label{tab:observations_parameters}}
\tablehead{
\colhead{Continuum} \vspace{-0.2cm}&
\colhead{Antennas} & 
\colhead{Date} & 
\colhead{On source} &
\colhead{Beam Size} & 
\colhead{P.A.} & 
\colhead{$\sigma$} & 
\colhead{$\Delta v$} & 
\colhead{S$_\nu$} \\
\colhead{Freq. (GHz)} \vspace{-0.2cm} & & & \colhead{ time (hrs)} & $^{\prime\prime \,\times \,\prime\prime}$& $^{\circ}$ & \colhead{(mJy beam$^{-1}$)} & \colhead{ (km s$^{-1}$)} & \colhead{ (mJy)} \\
\colhead{(1)}  & \colhead{(2)} & \colhead{(3)} & \colhead{(4)} & \colhead{(5)} & \colhead{(6)} & \colhead{(7)} & \colhead{(8)} & \colhead{(9)}
}
\startdata
\rule[-0.5ex]{0pt}{2.5ex}  221.1 & 6 & March 28, 2022 & 8 & 1.36$\times$1.23 & 85.9 & 0.40 & 1.22 & $2.72 \, \pm \, 0.42$ \\ 
\rule[-0.51ex]{0pt}{2.5ex}  271.2 & 6 & January 26, 2022 & 5 & 3.60$\times$2.67 & 9.4 & 0.53 & 1.22 & $2.03 \, \pm \, 0.54$ \\
\enddata
\tablecomments{SMA observational parameters for the continuum data. (1) Continuum frequency; (2) Number of antennas available during the observations; (3) Date of observation; (4) On source time; (5) Synthesized beam size and its (6) Position angle; (7) RMS noise in the map; (8) Velocity resolution; (9) Continuum flux density}
\end{deluxetable*}

\section{Data collection and analysis} \label{sec:3}

\subsection{SMA observations and data reduction}
\label{sec:3.1}

We present observations of 3C 303.1 (PI: Belinda Wilkes; 2021B-S006) obtained with two observing tracks on the SMA. In January 26, 2022, 3C 303.1 was observed using the compact (COM) configuration, targeting the $^{12}$CO(J=3-2) transition redshifted to 272.928 GHz (1.098 mm). The receivers RxB 240 and RxA 345 were used in a split-tuning setup. RxB 240 was tuned to local oscillator (LO) frequency of 261.206 GHz, covering a frequency range of 245.056-259.645 GHz as well as 265.056-279.642 GHz, while RxA 345 was tuned to LO frequency of 282.206 GHz, covering 265.056-279.642 GHz as well as 285.056-299.642 GHz. The $^{12}$CO(J=3-2) line was placed in the lower sideband (LSB) of RxA 345 (spectral window 3) and the upper sideband (USB) of RxB 240 (spectral window 4), both covering the range 271.068-273.355 GHz. In a follow-up observation in March 28, 2022, the continuum was observed at 221.1 GHz using an extended (EXT) configuration. Both receivers were tuned to a LO frequency of 225.541 GHz, covering 209.391-223.699 GHz as well as 229.391-243.699 GHz. Both observations used the SWARM correlator, which has 16,384 channels at a uniform resolution of 140 kHz ($\sim$ 0.15 km s$^{-1}$).

Both data sets were calibrated using the IDL-based software package called Millimeter Interferometer Reduction (MIR). The calibrated visibilities were averaged by a factor of 8 (corresponding to a resolution of 1.22 km s$^{-1}$), and then converted to the Common Astronomy Software Application (CASA) measurement set (MS) format for Imaging \footnote{We used the Python script \texttt{MIRFITStoCASA$\_$casa5} to convert from UVFITS format to CASA measurement set format. \url{https://lweb.cfa.harvard.edu/rtdc/SMAdata/format/convert/}}. The CASA task \texttt{tclean} was used for imaging. During interactive cleaning of dirty images, a mask was manually drawn in a region around the source. A threshold of three times the rms noise in the dirty image was used for interactive cleaning. Natural weighting was used to maximize the signal-to-noise (SNR) ratio. Multi-frequency synthesis (\texttt{mfs}) mode was used to create images. Image sizes of 256 $\times$ 256 and 512 $\times$ 512 pixels were used for the 221.1 GHz and 271.2 GHz maps, respectively. A cell size of 0.1 arcseconds was used for both maps. The continuum maps were then exported in FITS format and analyzed using the CASA-Viewer (see Fig.\,\ref{fig:continuum_maps}).

To create a spectral cube of $^{12}$CO(J=3-2), we first subtracted a continuum using the CASA task \texttt{uvcontsub$\_$old}, applying a first-order polynomial fit to line-free channels. We used $\pm100$ channels from the expected line position, at the observed frequency of 272.928 GHz. The image and cell size used were 256$\times$ 256 pixels and 0.1 arcseconds, respectively. We achieved an rms level (1$\sigma$) of 27.24 mK with a velocity resolution of 1.22 km s$^{-1}$. The resulting spectrum was analyzed using the Cube Analysis and Rendering Tool for Astronomy (CARTA) software.

\subsection{Archival SED data and Extinction corrections}
\label{sec:3.2}

We constructed the SED using archival photometric data from the literature and various data archives, covering frequencies from low-frequency radio to high-energy X-ray data. All data retrieved from data archives are retrieved with a search radius of 5 arcseconds. Some HST data are available, for example, the F330W filter \citep{Labiano2008a}. However, since the data is high-resolution and may miss low surface brightness features, we excluded UV data from HST. No Galaxy Evolution Explorer (GALEX) data are available for the source. Therefore, no UV data are included in our SED. The X-ray fluxes are consistent with one another; therefore, we only consider the X-ray flux from XMM-Newton as a total flux in  the SED fitting because of its broader bandpass. Similarly, since W3 and 12 $\mu\text{m}$ fluxes are very similar, we only use W3 flux in the SED fitting. All the photometric data are listed in Table \ref{tab:SED}.

\begin{deluxetable*}{cccc}

\tablecaption{Galactic and Internal dust reddening corrections \label{tab:extinction_corrections}}
\tablehead{
\colhead{Band/Filter} & \colhead{$\lambda$} & \colhead{Galactic dust extinction} & \colhead{Internal dust extinction} \\
\colhead{} & \colhead{(\AA)} & \colhead{$A(\lambda)_{\mathrm{Gal}}$ (mag)} & \colhead{$A(\lambda)_{\mathrm{H\alpha/H\beta}}$ (mag)}
}
\startdata
Ks & 21590 & 0.0112 & 0.1084 \\
F205W & 20636.08 & -- & 0.1166 \\
H & 16620 & 0.017 & 0.1652 \\
J & 12350 & 0.0275 & 0.2664 \\
F110W & 11233.62 & -- & 0.3103 \\
y & 9627.79 & 0.041 & 0.3978 \\
z & 8674.2 & 0.0488 & 0.4732 \\
i & 7534.96 & 0.0637 & 0.6179 \\
r & 6201.2 & 0.0835 & 0.8097 \\
g & 4849.11 & 0.1115 & 1.0819 \\
\enddata
\tablecomments{$\lambda$ is pivot wavelength taken from SVO Filter Profile Service \citep{SVOFilterService}. We use mean E(B-V) \cite{Schlafly2011} value of 0.0308 mag from the service maintained by IPAC. We assume a visual-to-dust-reddening ratio of $R_{v} = 3.1$ \citep{Cardelli1989, Fitzpatrick1999}. F(H$\alpha$)/F(H$\beta$) for 3C 303.1 is 3.85 \citep{Buttiglione2009}.}
\end{deluxetable*}

We corrected the photometric data for both Galactic and internal dust reddening. For Galactic dust reddening, we use \textit{Galactic Dust Reddening and Extinction} service maintained by IPAC at Caltech. We then calculated the extinction using the Python package \textit{extinction}, applying the \cite{Cardelli1989} extinction law. Internal dust reddening is computed using the Balmer decrement ratio (F(H$\alpha$)/F(H$\beta$)). The color excess corresponding to internal extinction is estimated as 

\begin{linenomath}
\[
E(B-V)_{\mathrm{H}\alpha/\mathrm{H}\beta} = \frac{2.5 \log(2.86/R_{\mathrm{obs}})}{k(\lambda_\alpha) - k(\lambda_\beta)}
\]
\end{linenomath}

where $R_{obs}$ is the observed Balmer decrement ratio (F(H$\alpha$)/F(H$\beta$)), $k(\lambda_\alpha)$ and $k(\lambda_\beta)$ are the extinction curves at $H_\alpha$ and $H_\beta$ wavelengths, respectively. The adopted values are $k(\lambda_\alpha)$ = 2.63 and $k(\lambda_\beta)$ = 3.71 \citep{Tremblay2010}. The derived Galactic and internal dust reddening magnitudes are listed in Table \ref{tab:extinction_corrections}. We also apply extinction corrections to the 2MASS and HST/NICMOS data as well as the Pan-STARRS data, since they exhibit significant internal dust reddening despite having minimal Galactic extinction. Only internal corrections are applied to the HST/NICMOS data, as Galactic extinction has already been accounted for.

\section{Results and analysis} \label{sec:4}

\subsection{Discussion on SMA observational results}
\label{sec:4.1}

\begin{figure*}[ht!]
\plotone{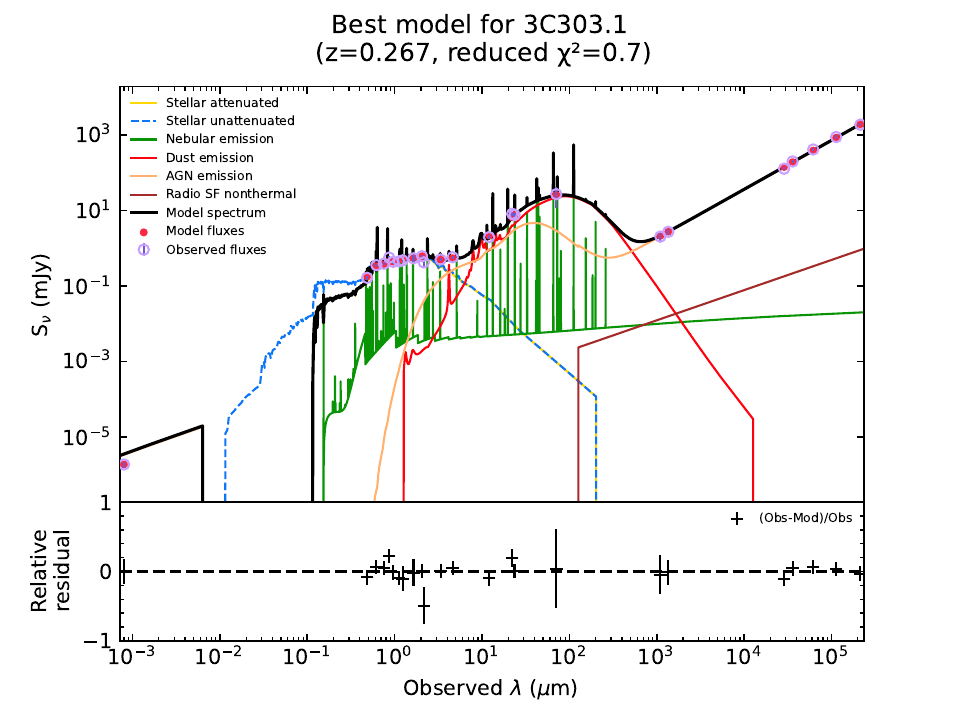}
\caption{The best SED model fit for 3C 303.1 using CIGALE. The reduced chi-square ($\chi^2$) value for our model SED fit is 0.7. The purple circles represent the observed photometry data with $1 \sigma$ error (vertical) bars (see Table \ref{tab:SED}). The red dots indicate the model fluxes. The solid black line shows the total model spectrum. The colored components represent different physical contributions: red for dust emission, orange for AGN emission, green for nebular emission, blue for unattenuated stellar emission, yellow for attenuated stellar emission (under stellar unattenuated and model spectrum), and brown for radio SF nonthermal emission. The bottom panel displays the residuals between the observed and model fluxes.}
\label{fig:sed_fit}
\end{figure*}

The continuum is detected in both data sets at signal-to-noise ratios (SNRs) of 6.80 at 221.1 GHz and of 3.83 at 271.2 GHz. The measured flux density is $2.72 \pm 0.42$ mJy at 221.1 GHz and $2.03 \pm 0.54$ mJy at 271.2 GHz (see Table \ref{tab:observations_parameters}). The flux density at 221.1 GHz was calculated by fitting a single Gaussian to the source, using a manually drawn region around the emission in the CASA-Viewer. At 271.2 GHz, since the source is unresolved, we used \texttt{IMSTAT} to estimate the flux density. For both flux densities, the background rms noise and 5\% flux scale uncertainty added in quadrature was adopted as the total uncertainty on the flux density. Self-calibration was attempted, but did not improve image quality due to the low SNR. From our data, the two-point spectral index ($\alpha$) between 221.1-271.2 GHz is $1.43 \pm 1.47$. These flux values lie along the extrapolated trend from the low-frequency radio data (see Fig.\,\ref{fig:sed_fit}). The integrated (1.4 - 271.2 GHz) $\alpha$ is $1.32 \pm 0.01$. The radio emission is consistent with the higher-frequency synchrotron emission. The previous measurement at 230 GHz with the IRAM 30-m telescope reported a flux density of $15 \pm 5$ mJy \citep{Steppe1995}, which lies significantly above the extrapolated low-frequency trend and is inconsistent with our measurements. Given that the continuum is consistent with the extrapolated synchrotron slope, we do not consider the IRAM 230 GHz measurement reliable for our analysis. The MERLIN image shows a clear double-lobed morphology (see Fig.\,\ref{fig:3C3303.1_OIII}; \cite{Ludke1998}). The Very long Baseline Interferometry (VLBI) observations of 3C 303.1 shows a compact feature that could correspond to the south-eastern lobe \citep{Popkov2021}. Both MERLIN and VLBI observations of 3C 303.1 are inconsistent with a beamed relativistic jet which would be the likely source of such strong variability. We therefore exclude the IRAM 230 GHz measurement from the SED (see Sec. \ref{sec:4.2}).

\begin{deluxetable*}{ccccccc}
\tablewidth{0pt}
\tablecaption{Observational parameters and results for $\mathrm{^{12}CO}~(J=3\text{--}2)$ \label{tab:observations_parameters_for_CO}}
\tablehead{
\colhead{$^{12}$CO(J=3-2) observed freq.} \vspace{-0.2cm}&  
\colhead{$\sigma$} & 
\colhead{$\Delta v$} &
\colhead{$\text{W}_{\text{CO}}$} & 
\colhead{$\text{N}_{\text{H}_2}$} &
\multicolumn{2}{c}{Molecular gas mass} \\
\colhead{(GHz)} \vspace{-0.2cm} & \colhead{(mK beam$^{-1}$)} & (km s$^{-1}$) & ($\text{K km s}^{-1}$) & ($\text{cm}^{-2}$) & \multicolumn{2}{c}{($\text{M}_\odot$)}\\
\colhead{(1)}  & \colhead{(2)} & \colhead{(3)} & \colhead{(4)} & \colhead{(5)} & \colhead{(6)} & \colhead{(7)}
}
\startdata
\rule[-0.5ex]{0pt}{2.5ex} 272.928 & 27.2 & 1.2 & $< 2.9$ & $< $8.2$ \times 10^{20} $ & $ < 1.7 \times 10^{9} $ & $ < 2.8 \times 10^{9} $
\enddata
\tablecomments{SMA observational parameters and results for \(^{12}\)CO(J=3-2) spectral line. (1) \(^{12}\)CO(J=3-2) observed frequency; (2) RMS noise in the \(^{12}\)CO(J=3-2) spectrum; (3) Velocity resolution; (4) Integrated \(^{12}\)CO(J=1-0) line intensity; (5) Column density of molecular hydrogen (\(N_{\mathrm{H}_2}\)); (6) Molecular gas mass derived from our \(^{12}\)CO(J=3-2) observations; (7) Molecular gas mass derived from another method (see Sec. \ref{sec:4.1})}
\end{deluxetable*}

No $^{12}$CO(J=3-2) line was detected at the expected position in our observations. No emission was detected out to approximately 0.5 arcmin from the galaxy at similar velocity. We therefore place upper limits on the molecular gas mass, calculated using two independent methods. In both cases, we estimate the upper limits on the molecular gas mass using the standard $^{12}$CO(J=1-0) calibration. To do this, we convert our $^{12}$CO(J=3-2) luminosity to an equivalent $^{12}$CO(J=1-0) luminosity using  the line ratio $r_{31} = L'_{\rm CO(3-2)}/L'_{\rm CO(1-0)}$. We adopt \(r\)$_{31}$ = 0.53 from \cite{Lamperti2020}, who derived this value for a sample of AGNs. The $3\sigma$ upper limit on the integrated \(^{12}\)CO(J=1-0) line intensity is given by the following formula: 

\begin{linenomath}
\[
W_{\mathrm{CO}} = \frac{3 \, \sigma \, V_{\mathrm{line}}}{r_{31}} \left( \frac{\Delta\nu}{V_{\mathrm{line}}} \right)^{1/2} \quad \mathrm{K\,km\,s}^{-1}
\]
\end{linenomath}

where \(\sigma\) is the RMS noise determined from line-free channels of 1D \(^{12}\)CO(J=3-2) spectrum, \(\Delta \nu\) is the velocity resolution, and \(V_{\mathrm{line}}\) is the assumed line width of 300 km s$^{-1}$ (assuming a rectangular profile for simplicity) \citep{ODea1993}. This is a good approximation because, for a Gaussian profile, \(V_{\mathrm{line}} \simeq 1.06 \times \mathrm{FWHM}\).

The empirical relation between the integrated $^{12}$CO(J=1-0) line intensity and the column density of molecular hydrogen (N$_{\mathrm{H}_2}$) can be expressed as follows:

\begin{linenomath}
\[
N_{\mathrm{H}_2} \simeq 2.8 \times 10^{20} \, W_{\mathrm{CO}} \, \mathrm{cm}^{-2}
\]
\end{linenomath}

This relation is calibrated for the Milky Way, under the assumption that the CO(J=1-0) line is optically thick \citep{Bloemen1986, Nishiyam2001}. However, we acknowledge that this conversion factor is subject to significant uncertainty.

The total mass of molecular hydrogen within the observing beam is obtained by spatially integrating over the Gaussian beam.

\begin{linenomath}
\[
M_{\mathrm{mol}} = \frac{\pi \, r^2}{4 \ln 2} \, N_{\mathrm{H}_2} \, m_{\mathrm{H}_2}
\]
\end{linenomath}

where \(m_{\mathrm{H}_2}\) is the mass of molecular hydrogen, and \(r\) is the full width at half maximum (FWHM) of the beam. A simplified version of this formula is as follows:

\begin{linenomath}
\[
M_{\mathrm{mol}} = 10.60 \, \theta^{2} \, D^{2} \, N_{\mathrm{H}_2} \, \mathrm{M_\odot}
\]
\end{linenomath}

where \(\theta\) is the beam FWHM in arcseconds, \(D\) is the luminosity distance in megaparsecs (Mpc) and \(N_{\mathrm{H}_2}\) is the column density in units of \(10^{20} \, \mathrm{cm}^{-2}\). The \(3\sigma\) upper limit on the molecular gas of 3C 303.1 is estimated to be \(1.7 \times 10^{9} \, \mathrm{M_\odot}\) (see Table \ref{tab:observations_parameters_for_CO}).

An alternative way of calculating the molecular gas mass using $^{12}$CO(J=1-0) luminosity is as follows: 

\begin{linenomath}
\[
L'_{\mathrm{CO}} = 23.5 \, \Omega_{\mathrm{s*b}} \, D_{\mathrm{L}}^2 \, W_{\mathrm{CO}} \, (1 + z)^{-3}
\]
\end{linenomath}

where \(L'_{\mathrm{CO}}\) is \(^{12}\)CO(J=1-0) line luminosity in K km s\(^{-1}\) pc\(^{2}\), \(\Omega_{\mathrm{s*b}}\) is the solid angle of the source convolved with the telescope beam in arcsec\(^2\), \(D_{\mathrm{L}}\) is the luminosity distance in Mpc, \(W_{\mathrm{CO}}\) is the observed integrated \(^{12}\)CO(J=1-0) line intensity in K km s\(^{-1}\) and \(z\) is the redshift of the source \citep{Solomon1997}. We use \(\Omega_{\mathrm{s*b}} \approx \Omega_{\mathrm{beam}}\) since the source is not resolved and is smaller than the beam. 

The molecular mass can be calculated from \(L'_{\mathrm{CO}}\) and the CO–H\(_2\) conversion factor as follows \citep{Audibert2022}: 

\begin{linenomath}
\[
M_{\mathrm{H}_2} = \alpha_{\mathrm{CO}} \, L'_{\mathrm{CO}}
\]
\end{linenomath}

We used a standard Galactic CO–H\(_2\) conversion factor, \(\alpha_{\mathrm{CO}} = 4.36 \, \mathrm{M_\odot}\, (\mathrm{K\,km\,s}^{-1}\mathrm{pc}^2)^{-1}\) \citep{Tacconi2013}. The \(3\sigma\) upper limit on the molecular gas is \(2.8 \times 10^{9} \, \mathrm{M_\odot}\). The average upper limit on the molecular gas mass, calculated in different methods, is \(2.3 \times 10^{9} \, \mathrm{M_\odot}\). This upper limit appears to be relatively low compared to the molecular gas masses compiled by \cite{2021A&ARv..29....3O}. Our estimate is below the median value of $\sim 6.8 \times 10^{9} \, \mathrm{M_\odot}$ of molecular gas masses compiled by \cite{2021A&ARv..29....3O}, but it does not lie at the extreme low end of $3 \times 10^{7} \, M_\odot$.

\begin{deluxetable*}{cccccc}
\tablewidth{0pt}
\tablecaption{3C 303.1 SED \label{tab:SED}}
\tablehead{
\colhead{Band/Filter} & 
\colhead{$\lambda_{\text{eff}}$ (\AA)} &
\colhead{Freq (Hz)} & 
\colhead{Flux (mJy)} & 
\colhead{Telescope/Instrument} &
\colhead{Reference}
}
\startdata
1.4 GHz & $2.14\times 10^{9}$ & $1.40\times 10^{9}$ & 1880.4 $\pm$ 66.2 & VLA & \cite{Condon1998} \\
2.65 GHz & $1.13\times 10^{9}$ & $2.65\times 10^{9}$ & 882 $\pm$ 19 & Effelesberg 100m & \cite{Mantovani2009} \\
4.85 GHz & $6.18\times 10^{9}$ & $4.85\times 10^{9}$ & 413 $\pm$ 8 & Effelesberg 100m & \cite{Mantovani2009} \\
8.35 GHz & $3.59\times 10^{8}$ & $8.35\times 10^{9}$ & 199 $\pm$ 4 & Effelesberg 100m & \cite{Mantovani2009} \\
10.45 GHz & $2.87\times 10^{8}$ & $1.05\times 10^{10}$ & 128 $\pm$ 3 & Effelesberg 100m & \cite{Mantovani2009} \\
221.1 GHz & $1.36\times 10^{7}$ & $2.21\times 10^{11}$ & 2.72 $\pm$ 0.42 & SMA & This work\\
271.2 GHz & $1.10\times 10^{7}$ & $2.71\times 10^{11}$ & 2.03 $\pm$ 0.54 & SMA & This work\\
70 $\mu\text{m}$ & $7.00\times 10^{5}$ & $4.28\times 10^{12}$ & 27 $\pm$ 15 & Spitzer/MIPS & \cite{Shi2005} \\
24 $\mu\text{m}$ & $2.40\times 10^{5}$ & $1.25\times 10^{13}$ & 7.6 $\pm$ 0.1 & Spitzer/MIPS & \cite{Shi2005} \\
W4 & $2.21\times 10^{5}$ & $1.36\times 10^{13}$ & 8.17 $\pm$ 0.67 & WISE & \cite{Wright2010} \\
12 $\mu\text{m}$ & $1.15\times 10^{5}$ & $2.60\times 10^{13}$ & 1.9 $\pm$ 0.2 & ISOCAM/ISO & \cite{Siebenmorgen2004} \\
W3 & $1.16\times 10^{5}$ & $2.59\times 10^{13}$ & 1.97 $\pm$ 0.09 & WISE & \cite{Wright2010} \\
W2 & 46028 & $6.51\times 10^{13}$ & 0.575 $\pm$ 0.015 & WISE & \cite{Wright2010} \\
W1 & 33526 & $8.94\times 10^{13}$ & 0.50 $\pm$ 0.01 & WISE & \cite{Wright2010} \\
Ks & 21590 & $1.39\times 10^{14}$ & 0.421 $\pm$ 0.105 & 2MASS & \cite{Skrutskie_2006} \\
F205W & 20636.08 & $1.45\times 10^{14}$ & 0.619 $\pm$ 0.006 & HST/NICMOS & \cite{deVries2000b} \\
H & 16620 & $1.80\times 10^{14}$ & 0.539 $\pm$ 0.088 & 2MASS & \cite{Skrutskie_2006} \\
J & 12350 & $2.43\times 10^{14}$ & 0.475 $\pm$ 0.071 & 2MASS & \cite{Skrutskie_2006} \\
F110W & 11233.62 & $2.67\times 10^{14}$ & 0.460 $\pm$ 0.004 & HST/NICMOS & \cite{deVries2000b} \\
y & 9627.79 & $3.11\times 10^{14}$ & 0.435 $\pm$ 0.017 & Pan-STARRS & \cite{Chambers2016} \\
z & 8674.2 & $3.46\times 10^{14}$ & 0.56 $\pm$ 0.01 & Pan-STARRS & \cite{Chambers2016} \\
i & 7534.96 & $3.98\times 10^{14}$ & 0.388 $\pm$ 0.008 & Pan-STARRS & \cite{Chambers2016} \\
r & 6201.2 & $4.83\times 10^{14}$ & 0.354 $\pm$ 0.003 & Pan-STARRS & \cite{Chambers2016} \\
g & 4849.11 & $6.18\times 10^{14}$ & 0.164 $\pm$ 0.007 & Pan-STARRS & \cite{Chambers2016} \\
0.5-7 keV & 3.31 & $9.07\times 10^{17}$ & $(1.74 \pm 0.52)\times 10^{-6}$ & Chandra & \cite{Massaro2010} \\
0.2-12 keV & 2.03 & $1.47\times 10^{18}$ & $(1.86 \pm0.27)\times 10^{-6}$ & XMM-Newton & \cite{Rosen2015} \\
\enddata
\tablecomments{$\lambda_{\text{eff}}$ (\AA)) is effective/pivot wavelength taken from SVO Filter Profile Service \citep{SVOFilterService}} for photometric data other than radio, sub-mm and X-ray. \\
\end{deluxetable*}

The null detection of \(^{12}\)CO(J=3-2) is an intriguing result. Strong radio-mode feedback signatures have been observed in 3C 303.1 (see Sec. \ref{sec:2}). The radio jets are driving warm gas outflows and generating shocks in the ISM of the host galaxy. These shocks may compress the gas and raise its temperature. This supports a scenario in which AGN-driven heating influences the state of the gas. When the gas is heated or ionized, \(^{12}\)CO(J=3-2) becomes an unreliable tracer of molecular gas \citep{Ginberg2011}. Such warm and dense molecular gas can instead be traced through higher-J transitions of $^{12}$CO, rotational and vibrational transitions of H\(_2\), \([\mathrm{C\,\textsc{ii}}]\,158\,\mu\mathrm{m}\) and \([\mathrm{C\,\textsc{i}}]\,609\,\mu\mathrm{m}\)
 \citep{Esposito2024, Allers2005, D'Eugenio2023}. It is also possible that clouds of the atomic and molecular gas are being displaced by the jets. The detection of warm gas outflows on a kpc scale further suggests that molecular gas may be expelled by the jets. The radio jets in compact radio sources have the potential to remove a significant fraction of the ISM from the host galaxy and, if coupling is efficient, could even expel the entire gas reservoir \citep{Nesvadba2007}.


\subsection{SED fitting using CIGALE} 
\label{sec:4.2}

Code Investigating GALaxy Emission (CIGALE) is a software package designed to model the SED of galaxies and estimate the physical properties of their host galaxies \citep{Boquien2019}. We used X-CIGALE (v2025), an extended version of CIGALE, to fit the SED from radio to X-ray regime \citep{Yang2022}. The code is widely used because it includes AGN emission templates. It models AGN emission including dust absorption and scattering, re-emission in the IR and X-ray emission. It accurately models SEDs of AGN-host galaxies and estimates both AGN and host galaxy physical properties. CIGALE reads observed photometric data and parameter specifications from a configuration file. Based on these input parameters, it generates a set of models and performs a Bayesian analysis of their likelihoods. It then outputs both Bayesian and best-fit parameter values, along with estimates of key physical properties. The Bayesian estimates are derived from the likelihood-weighted probability distribution of all models in the grid, and the associated uncertainties correspond to the standard deviation of these probability distributions \citep{Boquien2019}. All the input and output parameters are described in Table \ref{tab:CIGALE SED fit}.

The code first computes the star-formation history (SFH) of the host galaxy, then generates the intrinsic stellar unattenuated spectra using single stellar population (SSP) libraries. It then estimates nebular emission, both continuum and line, primarily originating from H\textsc{ii} regions in the galaxy. Next, it applies the effects of both dust attenuation and dust emission. It also models AGN and radio emission. Finally, it applies the redshift effect and absorption by inter-galactic medium (IGM). 

To compute SFH, we used delayed SFH with optional constant burst/quench (\texttt{sfhdelayedbq}) \citep{Ciesla2017}. This is an extension of the delayed SFH model (\texttt{sfhdelayed}). The \texttt{sfhdelayedbq} model accounts not only for the delayed SFH but also for recent variations, which may result from burst or quenching episodes, such as those that could be triggered by AGN feedback. The parameter $r_\mathrm{SFR}$, defined as the ratio of the SFR after/before burst/quench, reflects the recent variations. For $r_\mathrm{SFR}$, values both greater and less than 1 were provided as input, resulting in a best-fit value of 0.7. This suggests that star formation has been quenched after e-folding time of the main stellar population model ($\tau_\mathrm{main}$), possibly due to AGN-feedback. The stellar unattenuated spectrum is generated using simple stellar populations (SSPs). We used the widely adopted Bruzual \& Charlot 2003 (\texttt{bc03}) model. This model assumes a constant initial mass function (IMF) and metallicity to generate stellar spectra \citep{Bruzual2003}. We assume a Salpeter IMF in the mode to remain consistent with the different star formation indicators that also adopt a Salpeter IMF (see Sec. \ref{sec:4.3.2}). The Salpeter IMF overestimates the number of low-mass stars and hence provides conservative upper limits.

The nebular emission module accounts for both continuum and line emission from ionized gas, primarily from in H\textsc{ii} regions surrounding young, hot stars. For the electron density ($n_\mathrm{e}$), we use the default value of 100 $\mathrm{cm}^{-3}$. We also adopt the default value of $\textit{lines\_width} = 300~\mathrm{km\,s^{-1}}$, which is commonly observed in optical emission lines in CSS sources.

We adopt the modified \cite{Charlot2000} attenuation law (\texttt{dustatt\_modified\_CF00}) model. This model computes attenuation using two components: the birth clouds (BC) curve, which affects only young stars ($\leq10$ Myr) in the star-forming regions, and the diffused ISM component, which affects both young and old stars. Another available model is the modified attenuation law (\texttt{dustatt\_modified\_starburst}) which is specifically designed for the starburst galaxies \citep{Calzetti2000}. The dust emission is modeled using the \texttt{dl2014} \citep{Draine2014}. This model accounts for polycyclic aromatic hydrocarbons (PAHs), cold dust heated by a uniform and weaker radiation field intensity (Umin), and warm dust originating from photo-dissociation regions (PDRs), which is typically heated by intense radiation from young stars. Unlike the \texttt{Dale2014} \citep{Dale2014}, \texttt{dl2014} allows a broader variation of PAH fraction ($q_\mathrm{PAH}$), and a more flexible distribution of radiation field intensities (from $U_\mathrm{min}$ to $U_\mathrm{max}$).

We used the \texttt{skirtor2016} module to model AGN emission \citep{Stalevski2016}. The model assumes a two-phase torus structure (smooth and clumpy), accounting for dust absorption and scattering of AGN radiation, as well as re-emission in the IR by heated dust. Another available model, \texttt{fritz2006} \citep{Fritz2006}, does not account for the polar dust and therefore, underestimates MIR emission. The AGN fraction, denoted as frac$_{AGN}$, is defined as the ratio of AGN IR to the total IR luminosity of the host galaxy. We use the Gaskell law \citep{Gaskell2004}, which, unlike other available laws, treats AGN polar dust differently from ISM dust. The law is derived from AGN spectra and is therefore specific to AGN attenuation. The best-fit angle between the equatorial plane and torus edge degrees, also known as opening-angle is 50$^{\circ}$ and the viewing angle ($\theta$) is 40$^{\circ}$. Since the source is type 2, these angles lie at the boundary of type 1 and type 2 AGN. Inconsistencies may arise due to model limitations or anisotropic dust emission.

 We model radio synchrotron emission using parameters such as AGN radiation slope ($\alpha_{AGN}$), the FIR/radio correlation coefficient for star formation (q$_{FIR}$) and the radio-loudness parameter (R). The latter is defined as R = L$_{\nu, \,5GHz}$ / L$_{\nu, \,2500 \text{\AA}}$ where L$_{\nu, \,5GHz}$ is a 5 GHz luminosity and L$_{\nu, \,2500 \text{\AA}}$ is AGN intrinsic disk luminosity at 2500 $\text{\AA}$, at a viewing angle of 30$^\circ$. The X-ray emission is modeled using the approach described in \cite{Lopez2024}. This is the recent model, calibrated on modern AGN templates, which provides flexibility in cutoff, photon index and, most importantly in IR-X-ray scaling parameter $\alpha_{\mathrm{IRX}}$, defined as $\alpha_{\mathrm{IRX}} = \log (L_{\mathrm{int},\ 2\text{--}10\,\mathrm{keV}}/{L_{\mathrm{nuc},\ 12\,\mu\mathrm{m}}})$ where, $L_{\mathrm{int},\ 2\text{--}10\,\mathrm{keV}}$ is the intrinsic (absorption-corrected) X-ray luminosity in the 2–10 keV band, and $L_{\mathrm{nuc},\ 12\,\mu\mathrm{m}}$ is the nuclear mid-infrared luminosity at 12 $\mu \mathrm{m}$. It is difficult to constrain the X-ray parameters using only a single X-ray flux point, so the X-ray values used here are representative and chosen to be consistent with the overall SED.

\subsection{Discussion on SED fitting output parameters} \label{sec:4.3}

\begin{deluxetable*}{lr}
\tablewidth{0pt}
\tablecaption{Output parameters from SED fitting \label{tab:output_parameters}}
\tablehead{
\colhead{Parameter} & \colhead{value}
}
\startdata
Instantaneous SFR ($\text{M}_\odot$ yr$^{-1}$) & $23.8\pm1.2$  \\
Accretion power (erg s$^{-1}$) & $(1.5\pm0.1)\times10^{44} $ \\
Total AGN Luminosity (erg s$^{-1}$) & $(2.0\pm0.1)\times10^{44} $ \\
Total Dust Luminosity (erg s$^{-1}$) & $(4.7\pm0.2)\times10^{44} $ \\
Total Luminosity (erg s$^{-1}$) & $(1.3\pm0.6)\times10^{45} $ \\
Total Dust mass ($\text{M}_\odot$) & $(1.7\pm0.9)\times10^7$ \\
\enddata
\end{deluxetable*}

The key physical parameters derived from the best-fit SED model are listed in Table \ref{tab:output_parameters}. We focus on two physical parameters that play crucial role in galaxy evolution: Gas-to-dust mass ratio and star formation rate.

\subsubsection{Gas to dust mass ratio}
\label{sec:4.3.1}

The dust mass estimated from CIGALE is 1.7 $\times 10^7\, \mathrm{M_{\odot}}$. The upper limit on the gas to dust mass ratio (G/D) is 135. To calculate this ratio, we used only our upper limit on the molecular gas mass. This value is subject to significant uncertainties, primarily due to assumptions about the line ratio $^{12}$CO(J=3–2)/$^{12}$CO(J=1–0) (r$_{31}$) and the CO–H$_2$ conversion factor, both of which depend on local physical conditions. The 3$\sigma$ upper limit on the HI column density is 1.5 $\times 10^{20}\, \mathrm{cm}^{-2}$ \citep{Vermeulen2003}, which corresponds to a gas mass of 4.5 $\times 10^7 \,\mathrm{M_{\odot}}$. This value is significantly lower than the molecular gas mass of 2.3 $\times 10^9 \,\mathrm{M_{\odot}}$; therefore, we ignore the contributions from the atomic hydrogen when estimating the G/D mass ratio. Furthermore, a significant fraction of the molecular gas may be dark CO, i.e., H$_2$ that is present in the cloud not traced by CO emission, because CO molecules can be photodissociated in the outer, self-shielded layers of the cloud. This fraction can be substantial, typically $\sim30\%$ in Galactic molecular clouds \citep{Grenier2005}. There is also a possibility of heating or removal of cold gas by AGN feedback. Currently, no compiled list exists for dust masses in CSS/PS ratio for comparison. The dust masses for extended radio sources typically lie in the range of $10^8\, \mathrm{M_{\odot}}$  \citep{Podigachoski2015}. Our value of G/D ratio is at the lower end of typical Galactic values, which range from 120-180 \citep{Draine2007, Zubko2004}. Our upper limit on G/D is relatively lower than the median Galactic value of $\sim$161 \citep{Zubko2004}.

\subsubsection{SFR}
\label{sec:4.3.2}

We derive the SFR using the following methods:\\

\textit{Instantaneous SFR}: We estimate SFR using the \texttt{sfhdelayedbq} module in CIGALE, which, in addition to delayed SF, also accounts for recent quenching or burst events \citep{Ciesla2017}. It models the SFH as follows: 

\begin{linenomath}
\[
\text{SFR}(t) \propto
\left\{
\begin{array}{ll}
t e^{-t / \tau_{\mathrm{main}}} & \text{if } t \leq t_0 \\
r_{\mathrm{SFR}} \times \text{SFR}(t=t_0) & \text{if } t > t_0
\end{array}
\right.
\]
\end{linenomath}

\begin{deluxetable}{lr}
\tablecaption{Star Formation Rates (SFR) \label{tab:SFR}}
\tablehead{
\colhead{SFR Indicator} & \colhead{Value ($\mathrm{M_\odot}$~yr$^{-1}$ )}
}
\startdata
CIGALE & {$23.8\pm1.2$}\\
H$\alpha$ $^{a}$ & 20.5\\
$24 \,\mu \mathrm{m}$ $^{b}$ & 42.3\\
H$\alpha+24\, \mu \mathrm{m}$ $^{c}$  & 42.5\\
\enddata
\tablecomments{
$^{a}$ Corrected for internal dust attenuation (see Sec. \ref{sec:3.2}). \\
$^{b}$ Observed flux used to calculate SFR \\
$^{c}$ Not corrected for internal dust attenuation.}
\end{deluxetable}

where SFR(t) is SFR at time t, $\tau_{main}$ is the e-folding time of the main stellar population, t$_{0}$ is the time when a burst or quenching event occurs, and r$_{\text{SFR}}$ is the ratio of SFR after t$_{0}$ to SFR at t$_{0}$. \\

\textit{H$\alpha$ Recombination Line}: Young, hot stars primarily produce UV photons that ionize the surrounding nebular regions, also known as H\textsc{ii} regions. These H\textsc{ii} regions emit strongly in the optical regime, featuring prominent recombination and forbidden lines. Stars with masses greater than \(10\,\mathrm{M_\odot}\) and life times shorter than \(20\,\text{Myr}\) contribute significantly to these emissions. Thus, recombination and forbidden lines serve as effective tracers of such hot, young stars. The calibration by \cite{Kennicutt1998} relates SFR to the H$\alpha$ recombination line as follows:
\begin{linenomath}
\[
\text{SFR}\ (\mathrm{M_\odot}\ \text{yr}^{-1}) = 7.9 \times 10^{-42}\ L(\text{H}\alpha)\ (\text{erg\ s}^{-1})
\]
\end{linenomath}
 where $L(\text{H}\alpha)$ is corrected for both Galactic and internal dust reddening. This SFR is sensitive to short time scales, largely independent of prior SFH, and effectively traces an instantaneous SFR. It assumes the Salpeter IMF (0.1-100$\,\mathrm{M_\odot}$).\\

\textit{24~$\mu$m Mid-IR Continuum}: Optical and UV starlight is absorbed and re-emitted in the IR wavelengths, typically between 10 and 100 $\mu$m. The mid-IR continuum traces dust heated by young stars, providing an indirect SFR indicator. \cite{Rieke2009} calibrated the SFR using 24 $\mu$m MIPS flux as follows: 

\begin{align*}
\text{SFR}(\mathrm{M_\odot}\, \text{yr}^{-1}) &= 7.8 \times 10^{-10}\, L(24\,\mu\text{m}) \\
&\text{for } 6 \times 10^8\,L_\odot \leq L(24) \leq 1.3 \times 10^{10}\,L_\odot \\
\\
\text{SFR}(\mathrm{M_\odot}\, \text{yr}^{-1}) &= 7.8 \times 10^{-10}\, L(24\,\mu\text{m},\, L_\odot) \\
&\quad \times \left(7.76 \times 10^{-11}\, L(24\,\mu\text{m},\, L_\odot)\right)^{0.048} \\
&\text{for } L(24) > 1.3 \times 10^{10}\,L_\odot
\end{align*}

where $L(24~\mu\text{m})$ is the rest-frame luminosity measured with MIPS, without applying bandpass corrections.
High-luminoity CSS HERGs, such as 3C 303.1, often show reddened MIR spectra due to additional obscuration by dust.\\

\textit{Composite calibration of H$\alpha$ and 24~$\mu$m}: \cite{Kennicutt2009} developed a method that combines the observed emission of $H\alpha$ and 24 $\mu m$ to estimate the internal dust-attenuation corrected SFR. 

\begin{align*}
\text{SFR}(\mathrm{M_\odot}\, \text{yr}^{-1}) &= 7.9 \times 10^{-42} \Big[ L(\text{H}\alpha)_{\text{obs}} \\
&\quad +\; 0.020\,L(24\,\mu\text{m}) \Big]~(\text{erg\,s}^{-1})
\end{align*}

Here, both luminosities are not corrected for internal dust attenuation. \\

The SFRs derived from different indicators are listed in Table \ref{tab:SFR}. SFRs obtained using different methods are sensitive to the IMF and the SFH module used in CIGALE. SFRs derived from $H\alpha$ and 24~$\mu \mathrm{m}$ are likely contaminated by AGN contributions and should be interpreted as upper limits. In contrast, the SFR estimated using CIGALE is based solely on the modeled SFH and is free of AGN contamination. 3C 303.1 exhibits a moderate SFR, consistent with values typically observed in CSS and PS sources \citep{2021A&ARv..29....3O}.

\subsection{Connection to AGN feedback}
\label{sec:4.4}
Some of the best-fit results from CIGALE are $\mathrm{age}_\mathrm{main}$ = 4.5 Gyr, $\tau_\mathrm{main}$ = 4 Gyr and $\mathrm{age_{bq}} = $ 100 Myr with $r_\mathrm{SFR}$ = 0.7. This indicates that the quenching event occurred 100 Myr ($10^8$ yr) ago. The spectral and dynamical ages of 3C 303.1 are estimated to be of order $\sim10^5$ yr \citep{ODea2002}, which is significantly shorter than the quenching timescale inferred from the SED fitting. While these timescales are not directly comparable, several factors could explain this discrepancy.

First, the SED fitting performed with CIGALE does not include UV photometry, which is particularly sensitive to the star-formation history. Additional UV observations would therefore help to better constrain the star-formation history and the quenching timescale.

Second, if the spectral and dynamical ages are accurate, the currently observed radio activity may simply be too young to have caused the inferred quenching episode. In this scenario, earlier cycles of radio activity could have contributed to the suppression of star formation. Such episodic activity is often seen in double-double radio sources, where two or more pairs of radio lobes are observed due to earlier radio-jet activity \citep{Kaiser2000b, Saikia2009}. Evidence of previous radio activity is typically detected by faint, diffuse, low-frequency radio emission. Currently, 3C 303.1 remains unresolved in recent low-frequency observations from the third data release of the LOFAR Two-metre Sky Survey (LoTSS-DR3, \cite{Shimwell2026}) and future high-resolution observations at low-frequencies would be valuable for investigating earlier cycles of radio activity.

Third, the spectral and dynamical ages of compact radio sources can sometimes be underestimated. A class of objects known as frustrated radio sources remain compact because the radio jets interact strongly with a dense surrounding ISM \citep{2021A&ARv..29....3O}. In such cases, confinement of radio source can lead to an underestimation of the spectral and dynamic ages. However, given the $^{12}$CO(3-2) non-detection, this scenario is unlikely.

Finally, the quenching may not be driven solely by the radio source. Since the source is HERG, other forms of AGN feedback-such as AGN-driven winds or radiative feedback -may also contribute to the suppression of star-formation.

A similar discrepancy between radio source ages and star-formation quenching timescales has been discussed in \cite{Duggal2024}. We therefore suggest that earlier cycles of radio activity, or other forms of AGN feedback, may have contributed to the suppression of star formation, although the precise timing and dominant mechanism remain uncertain.

This declining SFR is consistent with the limited molecular gas reservoir implied by the $^{12}$CO(J=3--2) non-detection and the relatively low G/D ratio, suggesting that the cold gas supply has been reduced. Taken together, $^{12}$CO(J=3--2) non-detection, low G/D ratio, and the moderate but declining SFR point to the heating and/or removal of the cold gas by AGN feedback, which suppresses star formation. These results illustrate how AGN in 3C 303.1 regulate the cold gas content and influence the star formation activity.

\section{Conclusion} \label{sec:5}

The main results of our study are summarized below: 

\begin{itemize}
    \item We detected the continuum in our SMA observations, at 221.1 GHz and 271.2 GHz, with measured flux densities of $2.72 \, \pm \, 0.42$ mJy and $2.03 \, \pm \, 0.54$ mJy respectively. These values are consistent with extrapolations from low-frequency radio data, suggesting that the earlier IRAM 230 GHz flux measurement of $15 \pm 5$ mJy is inconsistent unless there is variability.
    \item We did not detect the $^{12}$CO(J=3--2) line. The average upper limit on the molecular gas mass estimated using two different methods, is $\sim 2.3 \times 10^{9} \, \text{M}_\odot$ (see Sec. \ref{sec:4.1}). This estimate is below the median value of $\sim 6.8 \times 10^{9} \, \mathrm{M_\odot}$ of molecular gas masses compiled by \cite{2021A&ARv..29....3O}, but it does not lie at the extreme low end of $3 \times 10^{7} \, \mathrm{M_\odot}$.
    \item Non-detection of $^{12}$CO(J=3--2) is an interesting result. 3C~303.1 exhibits strong radio-mode feedback signatures, including optical emission lines, UV and X-ray-emitting gas aligned with the radio jet axis. This indicates that outflows and shocks driven by radio jets in the ISM of the host galaxy are potentially disrupting the molecular gas. The molecular gas, particularly CO, may have been shock-heated and therefore not detected in the $^{12}$CO(J=3--2) line.
     \item Our upper limit on the gas-to-dust (G/D) mass ratio of 135, estimated using only molecular gas, lies at the end of typical Galactic values, which range from 120-180. The SFH derived from CIGALE suggests a declining SFR due to a recent quenching event ($\mathrm{age}_\mathrm{main}$ = 4.5 Gyr, $\tau_\mathrm{main}$ = 4 Gyr and $\mathrm{age_{bq}} = $ 100 Myr with $r_\mathrm{SFR}$ = 0.7). Additionally, SFRs derived from CIGALE ($23.8\pm1.2$ $\mathrm{M_\odot}$~yr$^{-1}$) and from several optical and IR indicators are moderate (ranging from 20.5 to 42.5 $\mathrm{M_\odot}$~yr$^{-1}$), consistent with the values observed in CSS/PS sources.
    \item Taken together, $^{12}$CO(J=3--2) non-detection, relatively low G/D ratio, and the moderate but declining SFR point to the heating and/or removal of the cold gas by AGN feedback, which suppresses star formation. These results illustrate how AGN in 3C 303.1 regulate the cold gas content and influence the star formation activity.

\end{itemize}

\begin{acknowledgments}
 
We thank an anonymous referee for helpful comments and suggestions, which improved the manuscript. R.B., C.O. and S.B. acknowledge the support of the Natural Sciences and Engineering Research Council (NSERC) of Canada. RB also acknowledges support from the University of Manitoba Graduate Enhancement of Tri-Council Stipends (GETS) program. BJW acknowledges the support of NASA Contract  NAS8-03060 (CXC). The Submillimeter Array is a joint project between the Smithsonian Astrophysical Observatory and the Academia Sinica Institute of Astronomy and Astrophysics and is funded by the Smithsonian Institution and the Academia Sinica. We recognize that Maunakea is a culturally important site for the indigenous Hawaiian people; we are privileged to study the cosmos from its summit. This research has made use of the NASA/IPAC Extragalactic Database, which is funded by the National Aeronautics and Space Administration and operated by the California Institute of Technology. This research has made use of the NASA/IPAC Infrared Science Archive, which is funded by the National Aeronautics and Space Administration and operated by the California Institute of Technology. This research has made use of the VizieR catalogue access tool, CDS, Strasbourg Astronomical Observatory, France (DOI : 10.26093/cds/vizier).

\end{acknowledgments}

\facilities{Submillimeter Array (SMA)}

\software{Common Astronomy Software Application \citep{CASA2022}, Cube Analysis and Rendering Tool for Astronomy \citep{CARTA2021}, Millimeter Interferometer Reduction (The MIR Cookbook\footnote{\url{https://lweb.cfa.harvard.edu/~cqi/mircook.html}}), Code Investigating GALaxy Emission \citep{Boquien2019}, \texttt{extinction} \citep{extinction2021}, Ned Wright’s Cosmological Calculator \citep{Wright2006}}

\begin{deluxetable*}{lcc}
\tabletypesize{\scriptsize}
\tablewidth{0pt}
\tablecaption{CIGALE SED Fitting parameters\label{tab:CIGALE SED fit}}
\tablehead{
\colhead{Parameter} & 
\colhead{Input values} &
\colhead{Best fit parameters}
}
\startdata
\multicolumn{3}{c}{SFH\texttt{(sfhdelayedbq)}} \\
e-folding time of the main stellar population model $\tau_\mathrm{main}$ (Gyr) & 3,4,4.5,5& 4\\
age of the main stellar population in the galaxy $\mathrm{age}_\mathrm{main}$ (Gyr) & 3,4,4.5,5,6 & 4.5\\
age of the burst/quench $\mathrm{age}_\mathrm{bq}$ (Myr) & 10,50,100,500,1000& 100\\
ratio of the SFR after/before burst/quench $r_\mathrm{SFR}$ (dimensionless) & 0.5, 0.7, 0.9, 1.1, 1.3 & 0.7\\
\multicolumn{3}{c}{SSP \texttt{(bc2003)}} \\
initial mass function $\mathrm{IMF}$ (0 = Salpeter, 1 = Chabrier) & 0 & 0 \\
metallicity (dimensionless) & 0.002, 0.008, 0.02, 0.05& 0.008\\
age of separation between young and old populations $\mathrm{separation\,age}$ (Myr) & 10, 50, 100& 10\\
\multicolumn{3}{c}{Nebular \texttt{(nebular)}} \\
ionisation parameter $\log U$ & -1, -2, -3& -2\\
gas metallicity $Z_\mathrm{gas}$ & 0.002, 0.011, 0.022, 0.041& 0.011\\
electron density $n_\mathrm{e}$ ($\mathrm{cm}^{-3}$) & 100 & 100\\
fraction of Lyman continuum photons escaping the galaxy $f_\mathrm{esc}$ & 0, 0.001, 0.005& 0\\
fraction of Lyman continuum photons absorbed by dust $f_\mathrm{dust}$ & 0, 0.001, 0.005& 0\\
\multicolumn{3}{c}{Dust attenuation \texttt{(dustatt\_modified\_CF00)}} \\
V-band attenuation in the interstellar medium $A_\mathrm{V,ISM}$ (mag) & 0.1, 0.15, 0.2, 0.3& 0.15\\
$\mu$ (dimensionless), ratio $A_\mathrm{V,ISM} / (A_\mathrm{V,BC} + A_\mathrm{V,ISM})$ & 0.3, 0.4, 0.5, 0.6& 0.3\\
power law slope of the attenuation in the ISM $\mathrm{slope}_{\mathrm{ISM}}$ & -0.5, -0.9, -1.5& -0.8\\
power law slope of the attenuation in the birth clouds $\mathrm{slope}_{\mathrm{BC}}$ & -2.5, -3, -3.5 & -3\\
\multicolumn{3}{c}{Dust emission \texttt{(dl2014)}} \\
mass fraction of PAH $q_\mathrm{PAH}$ (dimensionless) & 1, 1.5, 2, 5& 1.5\\
minimum radiation field $U_\mathrm{min}$ (Habing) & 1,5,7,10& 7\\
power-law slope $dU/dM \propto U^{\alpha}$ & 1.9, 2, 2.1, 2.2& 2.1\\
fraction illuminated from $U_\mathrm{min}$ to $U_\mathrm{max}$ $\gamma$ (dimensionless) & 0.7,0.8,0.9,0.95& 0.8\\
\multicolumn{3}{c}{AGN \texttt{(skirtor2016)}} \\
torus optical depth at 9.7 $\mu m$ $\tau_{9.7}$ & 3,5,7,9,11& 9\\
torus density radial parameter $pl$ ($\rho \propto r^{-p} e^{-q |\cos(\theta)|}$) & 0,0.5,1,1.5& 1\\
torus density angular parameter $q$ ($\rho \propto r^{-p} e^{-q |\cos(\theta)|}$) & 0,0.5,1,1.5& 0.5\\
angle between equatorial plane and torus edge (degrees) oa& 0,10,20,30\ldots80& 50\\
ratio of maximum to minimum torus radii R & 10,20,30& 30\\
viewing angle $\theta$ (degrees) & 0, 10,20,30\ldots90& 40\\
AGN fraction in total IR luminosity $\mathrm{fracAGN}$ & 0.1,0.2,0.3,0.4,0.5& 0.3\\
extinction law of polar dust (SMC) & 2 (\cite{Gaskell2004})& 2\\
$E(B-V)$ of polar dust & 2,2.3,2.5,2.7& 2.6\\
temperature of polar dust (K) & 75,100,125,150,175& 150\\
emissivity of polar dust & 0.3,0.4,0.5,0.6& 0.5\\
\multicolumn{3}{c}{Radio \texttt{(radio)}} \\
FIR/radio correlation coefficient for star formation $q_\mathrm{IR, SF}$ & 1.8 & 1.8\\
slope of power-law synchrotron emission for star formation $\alpha_\mathrm{SF}$ & 0.8 & 0.8\\
radio-loudness parameter for AGN $R_\mathrm{AGN}$ & $10^3,10^4,10^5,10^6$& $10^5$\\
slope of AGN radio emission $\alpha_\mathrm{AGN}$ & 1.2,1.3,1.4& 1.3\\
\multicolumn{3}{c}{X-ray \texttt{(lopez2024)}} \\
AGN photon index $\Gamma$ & 1.8,1.7,1.6& 1.8\\
exponential cutoff energy of the AGN spectrum $E_\mathrm{cut}$ (keV) & 30,40,50,100& 40\\
maximum deviation from $\alpha_\mathrm{ox}-L_{2500\,\text{\AA}}$ relation & 1,1.05,1.1,1.2,1.3& 1.05\\
\enddata
\tablecomments{The parameters not mentioned in the table were set to their default values.}
\end{deluxetable*}

\clearpage

\bibliography{sample7}{}
\bibliographystyle{aasjournalv7}

\end{document}